\documentclass[vecphys]{svmult}

\usepackage{makeidx}         
\usepackage{graphicx}        
\usepackage{multicol}        
\usepackage{cite}            
\usepackage{latexsym}
\usepackage{amssymb}
\usepackage{amsfonts}
\usepackage{amsmath}                             
\usepackage[bottom]{footmisc}

\makeindex             

\usepackage{soul}
\usepackage{color}

\begin{document}

\title{Fano Resonances in Flat Band Networks}
\author{Ajith Ramachandran \and Carlo Danieli  \and Sergej Flach}
\institute{Center for Theoretical Physics of Complex Systems, Institute for Basic Science, Daejeon, South Korea
}

\maketitle

\begin{abstract}
Linear wave equations on Hamiltonian lattices with translational invariance are characterized by an eigenvalue band structure in reciprocal space. 
Flat band lattices have at least one of the bands completely dispersionless. Such bands are coined flat bands. Flat bands occur in fine-tuned networks,
and can be protected by (e.g. chiral) symmetries.  Recently a number of such systems were realized in structured optical systems, exciton-polariton condensates, 
and ultracold atomic gases. Flat band networks support compact localized modes. Local defects couple these compact modes to 
dispersive states and generate Fano resonances in the wave propagation. Disorder (i.e. a finite density of defects) leads to
a dense set of Fano defects, and to novel scaling laws in the localization length of disordered dispersive states. Nonlinearities can preserve the compactness of flat band modes, along with 
renormalizing (tuning) their frequencies. These strictly compact nonlinear excitations induce tunable Fano resonances in the wave propagation of a nonlinear flat band lattice.
\end{abstract}


\section{Introduction}
\label{sec:intro}

In this chapter we will discuss Fano resonances induced by defects, disorder and nonlinearities in flat band networks.  More specifically, we will
present phenomena of resonant scattering occurring in lattices that, in the crystalline case, exhibit the existence of one (or more) {\it dispersionless} (or {\it flat}) band.

One of the main reasons to study the class of flat band networks is the existence of compact localized states (CLS), flatband eigenstates of the system which extend over a strictly finite number of lattice sites. 
Differently from Anderson localization, where localized states may exist due to uncorrelated disorder, CLS appear in ordered systems, and their existence
is protected by local symmetries which induce destructive interference in the lattice that suppress the propagation out of the compact domain of their nonzero amplitudes.

Introduced around the late 1980s, this class of models recently shifted into the focus of interest of a broad community due to mathematical advancements
as well as experimental realizations. Indeed, compact localized states are optimal candidates for transmission in networks of photonic waveguides that minimize
the diffraction due to the destructive interference. Furthermore, flat band networks and the observation of compact localized states have been experimentally
realized with exciton-polariton and Bose-Einstein condensate.

In several cases, the complete set of compact localized states can be fully detangled from the dispersive bands through suitable unitary transformations. 
This mathematical procedure allowed to extensively study several physical effects. In the following chapter, we will discuss Fano resonances induced by impurities,
onsite perturbations and nonlinear terms in flat band networks. We will discuss our findings using one of the simplest and most re-known flat band networks -
the {\it cross-stitch} lattice - as a test bed to present our results.

This chapter is structured in the following way. In the introduction we will review basic concepts of Fano resonances and of flat-band lattices.
We will then discuss how single impurities can induce Fano resonance in the system. This is further discussed in the following section, where we
discuss absence of transport in the flat band lattice due to the presence of uncorrelated disorder and quasiperiodic potentials. At last, we present Fano
resonances induced in a perfectly periodic flat band structure in the presence of  additional nonlinear terms.

\subsection{ Fano resonances}

In the quantum mechanical study of auto-ionising atoms, Ugo Fano introduced a new type of resonance mechanism
to explain the asymmetric profile of spectral lines \cite{fano1935sullo,fano1961effects}. 
The microscopic origin of the asymmetric line profile is due to constructive and destructive interference of the light continuum of states with a localized state hosted by the atom,
giving rise to additional paths for an incoming wave to scatter \cite{miroshnichenko2010fano}.
The resulting constructive or destructive interference gives rise to either perfect transmission or complete reflection. 
Fano derived this line shape as \cite{fano1961effects}
\begin{equation}
 T(s)=\frac{(s+q)^2}{s^2+1}\;,
 \label{Fano}
\end{equation}
where $s=(E-E_R)/(\Gamma/2)$ and $E_R$ is the resonance energy, $\Gamma$ the line width and $q$ the asymmetry parameter.
Recently, Fano resonances were observed in a variety of cases such as 
electronic transport in quantum dots \cite{johnson2005charge,gores2000fano,bulka2001fano,torio2004spin}, wires and tunnel junctions \cite{franco2003fano}, 
Mie and Bragg scattering in photonic crystals \cite{rybin2010bragg,rybin2009fano}, and
bilayer graphene nano-structures \cite{mukhopadhyay2011signature}.

The Fano-Anderson model is one of the simplest models that describes the physics and main features of Fano resonances \cite{miroshnichenko2010fano}.
This model consists of a tight-binding chain with nearest-neighbor hopping,
and a side-coupled discrete defect state. 
The Fano-Anderson Hamiltonian is given by
\begin{equation} 
H=C\sum_n(\psi_n\psi_{n-1}^*+c.c)+E_F|\phi|^2+V_F(\phi^*\psi_0+c.c.)\;.
\label{eq:Fano_Anderson_eq0} 
\end{equation}
In the absence of the coupling $V_F=0$ it supports propagation of plane waves with dispersion relation 
$\omega_k=2C\cos{k}$, while the isolated defect state has energy $E_F$. For nonzero coupling $V_F \neq 0$
equations of motion read
\begin{equation}
\begin{split}
i \dot{\psi}_n  &=  C\big( \psi_{n+1} + \psi_{n-1}\big) + \delta_{n,0} V_F\phi \;,   \\
i \dot{\phi}  &= E_F\phi + V_F\psi_0\ .
\end{split}
\label{eq:Fano_Anderson_eq}
\end{equation}
\begin{figure}\begin{center}
\includegraphics[width=0.8\textwidth ]{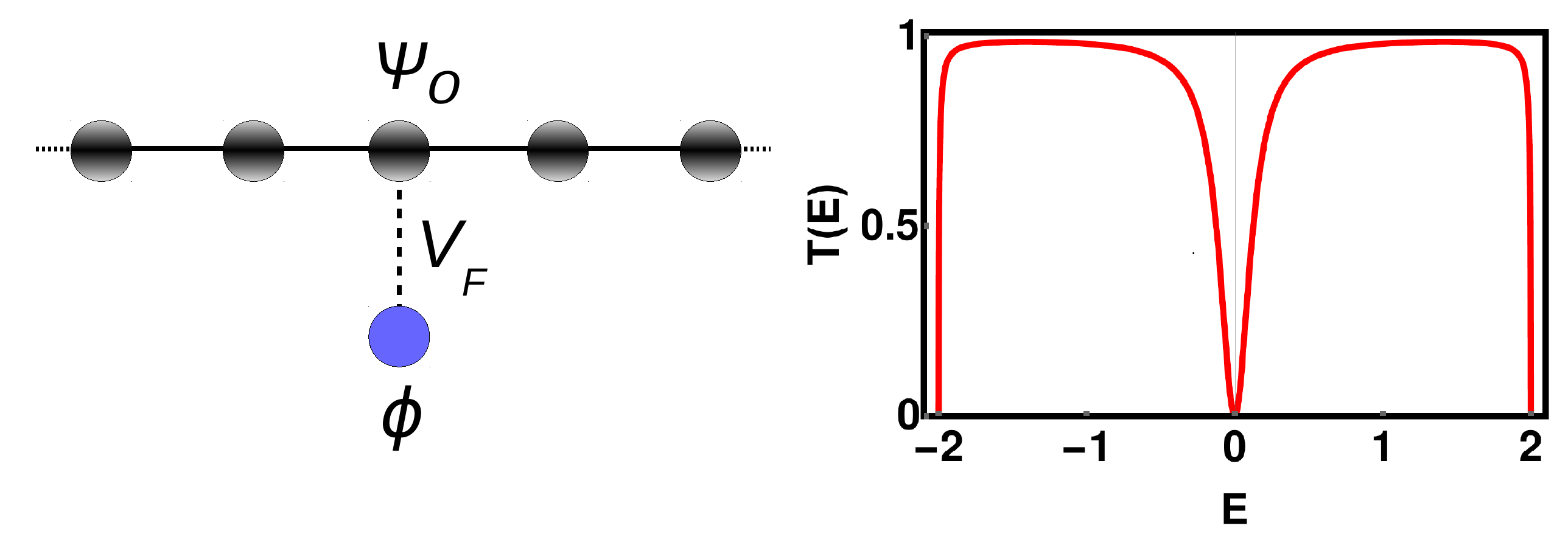}  \\
\end{center}
\caption{Schematic view of the Fano-Anderson model (left). Transmission coefficient for the Fano-Anderson model (right). 
Here: $C=1$, $V_F=0.5$ and $E_F=0$.
} \label{f2}
\end{figure}
A traveling wave has now the choice of two scattering channels: it can either bypass the defect state, or populate the state and return back to the chain.
The existence of these two paths gives rise to destructive interference, and a Fano resonance. To the left and right of the defect, we write the propagating modes in the usual scattering formulation
\begin{equation}
\psi_n =
\begin{cases}
& \tau e^{kn} + \sigma e^{- i k n} \ ,\quad n<0 \\
& \rho e^{ikn}\ ,\qquad\qquad\ \    n\geq 0
\end{cases}\ ;
\label{eq:CS_ext_wave}
\end{equation}
where $\tau$, $\sigma$ and $\rho$ are the incoming, reflected, and transmitted amplitudes respectively. 
The transmission coefficient $T(\omega) = |\rho / \tau |^2$ can be obtained using the transfer matrix method \cite{tong1999wave}:
\begin{equation}
\begin{split}
T(\omega_k) = \frac{\alpha_k^2}{1+\alpha_k^2}\;,
\end{split}
\label{eq:Fano_Anderson_FR}
\end{equation}
where
\begin{equation}
\begin{split}
\alpha_k = c_k\frac{E_F - \omega_k}{V_F^2}\ ,\qquad c_k=2C\sin k\;.
\end{split}
\label{eq:Fano_Anderson_cond2}
\end{equation}
For the resonant frequency $\omega_k = E_F$, the scattering along the two channels generates destructive interference leading
to a complete suppression of the wave transmission. 
The resonance width is proportional to the squared 
coupling strength $V_F^2$.
The asymmetry parameter of the Fano resonance for the Fano-Anderson model vanishes $q=0$ \cite{miroshnichenko2010fano}, thus the transmission profile is symmetric.

\subsection{ Flat band networks}\label{sec:FB}

Flat band networks are translationally invariant tight binding lattices (also coined continuous time quantum walks) of various dimension and hopping range which support at least one dispersion-less band in the energy spectrum \cite{flach2014detangling}.
In this chapter, we will focus on the case of a one-dimensional lattice modeled with nearest-neighbor hopping between unit cells:
\begin{equation}
i \dot{\vec{\psi}}_n = \epsilon_n \vec{\psi}_n + H_0\psi_n + H_1\vec{\psi}_{n+1} + H_1^\dagger\vec{\psi}_{n-1}\ .
\label{eq:FB_ham1}
\end{equation}
Here $\vec{\psi}_n = (\psi_n^1,\dots,\psi_n^\nu)^T\in \mathbb{C}^\nu$ is a wave function vector with $\nu$ complex scalar components
residing in the $n$th unit cell, and $H_0,H_1\in M_\nu(R)$ are $\nu\times\nu$ square matrices representing intra-cell and nearest neighbor inter-cell hoppings, respectively. 
The optional onsite perturbation $\epsilon_n$ of Eq.(\ref{eq:FB_ham1}) is given by a diagonal square matrix
$ \epsilon_n= \mbox{diag}\left( \epsilon_n^a, \epsilon_n^b, \dots, \epsilon_n^\nu \right)$, where $ \epsilon_n^i $ 
are so-called on-site energies originating from some external potential (field).
Using the ansatz $\psi_n = A_n e^{-i E t}$, the eigenvalue problem reduces to
\begin{equation}
  E\vec{A}_n = \epsilon_n \vec{A}_n + H_0 \vec{A}_n + H_1 \vec{A}_{n+1} + H_1^\dagger \vec{A}_{n-1}\ .
\label{eq:FB_ham_EP}
\end{equation}
For vanishing onsite energies $ \epsilon_n = 0$ the equations are invariant under discrete lattice translations, and the Bloch theorem leads to the ansatz $\vec{A}_n  = e^{i {\bf k} n } \vec{\varphi}_{\bf k}$
and a Bloch Hamiltonian $H({\bf k})$:
\begin{equation}
\quad  E \vec{\varphi}_{\bf k}  =  H({\bf k}) \vec{\varphi}_{\bf k}  \equiv \big(H_0 +  e^{i {\bf k}  }H_1 +  e^{ -i {\bf k}  } H_1^\dagger \big) \vec{\varphi}_{\bf k} \ .
\label{eq:FB_Bloch_mat}
\end{equation}
Solving the eigenvalue problem for $H({\bf k})$ we arrive at 
the band structure with $\nu$ $\bf k$-periodic bands $E_{1,2,...,\nu}({\bf k})$ and the corresponding set of polarization eigenvectors.

\begin{figure}[h]
 \centering
 \includegraphics[ width=0.65\columnwidth]{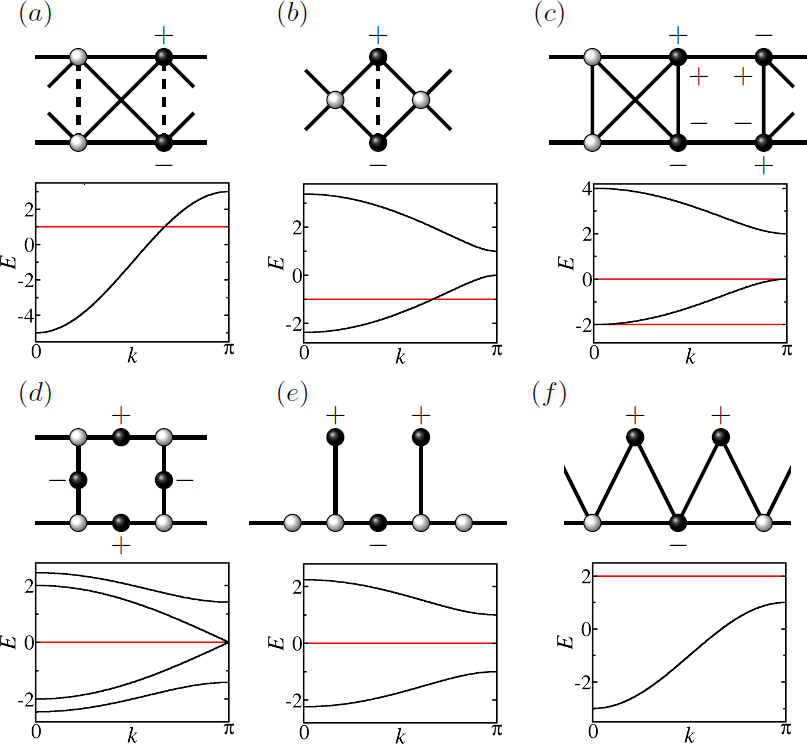}
 \caption{One-dimensional flat band topologies. 
(a) cross-stitch, $\kappa=1$, $U=1$; (b) diamond, $\kappa=1$, $U=1$; (c) one-dimensional pyrochlore, $U=1$; (d) one-dimensional Lieb $U=2$; (e) stub. $U=2$; (f) saw-tooth, $U=2$. Figure taken from \cite{flach2014detangling}.}
 \label{fig:FB_top}
\end{figure}

For flat band lattices at least one of the energies $E_m({\bf k})= const$ resulting in macroscopic degeneracy.
Relevant perturbations can lift the degeneracy and qualitatively change the nature
of the eigenstates \cite{perchikov2017flat,huber2010bose,aoki1996hofstadter,Leykam2013Flat,leykam2017localization,bodyfelt2014flatbands,danieli2015flatband}.
Due to the degeneracy, Bloch states of the flat band can be superimposed still yielding a valid eigenstate. It turns out,
that in many cases superpositions exist which yield not only localized, but even compact localized states (CLS).
The origin of these compact localized states is destructive interference that prevents diffraction and effectively decouples them from rest of the lattice.
The set of CLS can be orthogonal and linearly independent, and nonorthogonal but still linearly independent.
In those cases the CLS set spans the entire Hilbert subspace of the flat band, and there exists some unitary transformation
which connects the CLS set with the corresponding Bloch eigenstate set. In dimension $d \geq 2$, and in the presence
of band touchings of the flat band and a dispersive band, it is also possible that the CLS set is linearly dependent and 
incomplete. However, gapping the flat band away from the dispersive spectrum, or in general in one dimension,
linear dependence can be avoided.
In one dimension, the  CLS can be expressed in the following form:
\begin{equation}
 \vec{\psi}_{n_0} (t) = \Bigg\{ \sum_{l=0}^{U-1} \Bigg[  \sum_{j=1}^{\nu} a_{l,j} A_{l,j} {\bf e}_j \Bigg] \delta_{n,n_0 + l}\Bigg\} e^{-i \Omega t}  \ .
\label{eq:FB_states1}
\end{equation}
where $\Omega = E_{FB} $. 
Here ${\bf e}_j $ are the basis vectors in $\mathbb{R}^\nu$, and the integers 
$a_{i,j} \in \{ 0,\pm 1 \}$ pin down the locations of the nonzero CLS amplitudes
$A_{i,j}$.
The integer number $U$ counts the number of unit cells occupied by one CLS \cite{flach2014detangling}. 
This number is also called the {\it class} of the CLS. 
If the class $U$ flat band network is smoothly modified such that the CLS turn linearly dependent, the consequence
is that the class $U$ is reduced. The class $U=1$ always possesses an orthogonal and linearly independent CLS set.
In Fig.\ref{fig:FB_top} we show some examples of flat band networks.

Flat band networks are discrete geometrical structures that find applications in
distortion free image transmission through photonic lattices \cite{vicencio2015observation, weimann2016transport}, artificial quantum dot arrays \cite{brandes2005coherent}, ultracold atoms \cite{taie2015coherent}, 
 microwaves in dielectric resonator networks \cite{bellec2013tight,casteels2016probing},
 light-matter exciton-polariton condensates \cite{masumoto2012exciton}, among others. Furthermore, compact localized states
were experimentally observed in photonic lattices \cite{vicencio2015observation}, structured microcavities for exciton-polariton condensates
\cite{masumoto2012exciton}, and electronic circuits \cite{qiu2016designing}.
%

Attempts to construct flat band generators were based on graph theory \cite{mielke1991ferromagnetism}, local cell construction \cite{tasaki1992ferromagnetism}, Origami rules in decorated lattices \cite{dias2015origami},
and repetitions of mini arrays \cite{morales2016simple}. All these generators are focussing on subclasses of flat band networks
with some additional symmetry or property. The classification via CLS properties including the class $U$ is in principle complete
\cite{flach2014detangling}. The most general flat band generator for $U=1$ was obtained in Ref. \cite{flach2014detangling}.
The extension to $U=2$ proved already to be more involved, but was completed by 
Walaiymu {\it et.al.} \cite{maimaiti2017compact} 
for one-dimensional networks with two bands (one dispersive, one flat) and nearest neighbour hoppings.
A chiral flat band network generator for bipartite networks with majority sub-lattices 
was obtained in Ref.  \cite{Ramachandran2017chiral}, which yields chiral symmetry protected flat bands in any space dimension,
and no further restrictions - 
even the loss of translational invariance is not destroying the macroscopic degeneracy in the energy spectrum and the correspondent CLS.
Properly tuned magnetic fields can yield all bands flat \cite{vidal2000interaction}.
Further results concern non-Hermitian flat band networks \cite{Leykam2017Flat}, Bloch oscillations \cite{khomeriki2016landau},
topological flat Wannier-Stark bands \cite{kolovsky2017topological}, and the existence of nontrivial superfluid weights \cite{peotta2015superfluidity}, among others.
%

%
%
Let us consider the most simple and generic flat band case of $U=1$, and
of one dispersive and one flat band.
This cross-stitch lattice has two spectral bands and is obtained with the following matrices in Eq.(\ref{eq:FB_ham1})
(see Fig.\ref{fig:FB_top}(a)):
\begin{equation}
H_0 = \left(
\begin{array}{ccc}
0 & \kappa \\
\kappa & 0  
\end{array} \right),\quad
H_1 = \left(
\begin{array}{ccc}
1 & 1 \\
1 & 1  
\end{array} \right)\;.
\label{eq:CSgeo}
\end{equation}
The wave equations read  
\begin{equation}
\begin{split}
i\dot{\psi}^1_n &= \epsilon_n^a \psi^1_n - \psi^1_{n-1} - \psi^1_{n+1} - \psi^2_{n-1} - \psi^2_{n+1} - \kappa \psi^2_{n} \;, \\
i\dot{\psi}^2_n &= \epsilon_n^b \psi^2_n - \psi^2_{n-1} - \psi^2_{n+1} - \psi^1_{n-1} - \psi^1_{n+1} - \kappa \psi^1_{n} \;.
\end{split}
\label{eq:CS_lin1}
\end{equation}
For the dispersive band $ E_{DB}(k) = -\kappa - 4\cos(k)$,  and for the flat band $E_{FB} = \kappa $. 
The hopping strength $\kappa$ tunes the flat band energy and the relative position of the two bands, which can overlap for $|\kappa| \leq 2$, or can be gapped otherwise.
The compact localized states (\ref{eq:FB_states1}) are given by
\begin{equation}
\begin{split}
\vec{\psi}_{n,n_0} = 
 A \left(
\begin{array}{ccc}
1 \\
-1   
\end{array} \right)
 \delta_{n,n_0} e^{-iE_{FB}t}    
\end{split} \;.
\label{eq:CS_CLS}
\end{equation}
The detangling procedure is a unitary transformation which is applied to the vector space of each unit cell: 
\begin{equation}
\left(
\begin{array}{ccc}
p_n \\
f_n  
\end{array} \right) = D \vec{\psi}_n ,\quad  D = 
\frac{1}{\sqrt{2}}\left(
\begin{array}{ccc}
1 & 1 \\
1 & -1  
\end{array} \right)\ ,
\qquad 
\epsilon_n^{\pm}=(\epsilon_n^a \pm \epsilon_n^b)/2.
\label{eq:fano-CS}
\end{equation}
This yields 
\begin{equation}
\begin{aligned}
(E+\kappa)\, p_n &= \epsilon_n^+\, p_n + \epsilon_n^-\, f_n - 2\left( p_{n-1} + p_{n+1} \right), \\
(E-\kappa)\, f_n &= \epsilon_n^+\, f_n + \epsilon_n^-\, p_n\;.
\end{aligned}
\label{eq:CS_rotated}
\end{equation}
%
%
%
In the ordered case $\epsilon_n^i = 0 $ the flat band states $f_n$ are decoupled from the dispersive ones $p_n$.
Onsite perturbations introduce non-zero couplings $\epsilon_n^\pm \neq 0 $ in Eq.(\ref{eq:CS_rotated}) which hybridize the two families of states. 
%

\section{Single Local Defects}\label{sec:local_def}
\label{sec:defect}

In the presence of a flat band, an impurity locally hybridizes one or few renormalized CLS of the flat band with the dispersive bands,
turning them into a Fano state
which can lead to a Fano resonance.

\begin{figure}\begin{center}
\includegraphics[width=0.7\textwidth ]{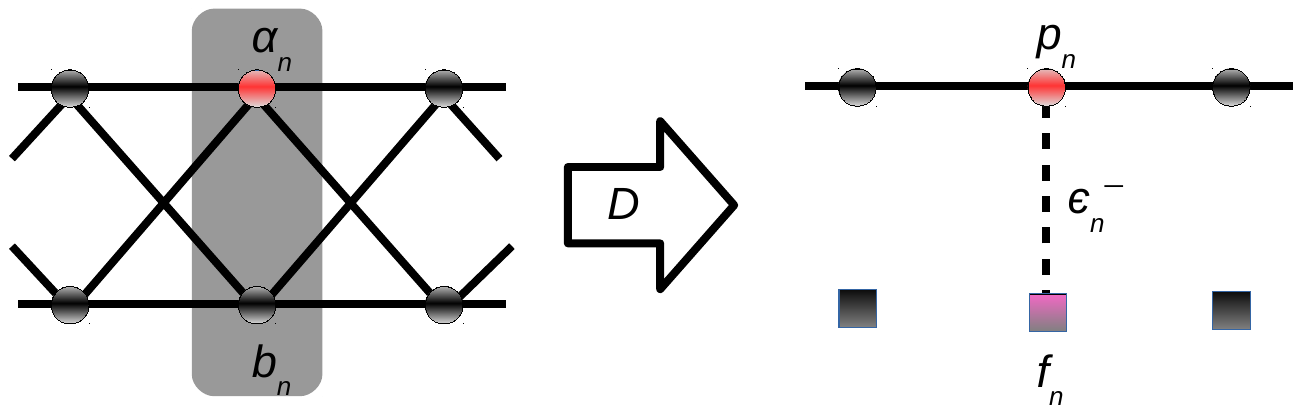}  \\
\end{center}
\caption{The cross-stitch lattice (left) and the detangled form (right). Here, the red dot on the cross-stitch lattice represents a local on-site defect. In the detangled
chain, a localized state $f_n$ is coupled to the linear chain due to the presence of defect.} \label{f2a}
\end{figure}

Consider an onsite energy variation at unit cell $n_0$ which
can be expressed as $\epsilon_n^{a,b}=\epsilon_n^{a,b}\delta_{n,n0}$.
It follows
\begin{align}
&Ep_n=\epsilon_n^+ \delta_{n,n0}p_n+\epsilon_n^-\delta_{n,n0}f_n-2(p_{n-1}+p_{n+1}) \;, \\
&Ef_n=\epsilon_n^+\delta_{n,n0}f_n+\epsilon_n^-\delta_{n,n0}p_n \;.
\end{align}
The resulting generalized Fano-Anderson model is shown in Fig. \ref{f2a}. Precisely one of the CLS is renormalized, and coupled to the dispersive chain which in addition
is perturbed by a simple defect at the site which is coupled to the CLS.

Excluding $f_{n0}$ the dispersive wave equation is reduced to 
\begin{equation}
E p_n=\left[\epsilon_n^+ + \frac{(\epsilon_n^-)^2}{E-\epsilon_n^+}\right]\delta_{n,n_0}p_n-2(p_{n-1}+p_{n+1}) \;.
\end{equation}
We then obtain the transmission coefficient as
%
%
%
\begin{equation}
 T(E)= \frac{16-\text{E}^2}{ 16-\text{E}^2+\frac{\left[2 \epsilon_n^+ (\text{E}-\epsilon_n^+)+(\epsilon_n^-)^2\right]^2}{ (\text{E}-\epsilon_n^+)^2}} \;.
\end{equation}
A Fano resonance appears at $E=\epsilon_n^+$. Some scattering outcomes are plotted in Fig.\ref{f2}. 
\begin{figure}\begin{center}
\includegraphics[width=0.7\textwidth ]{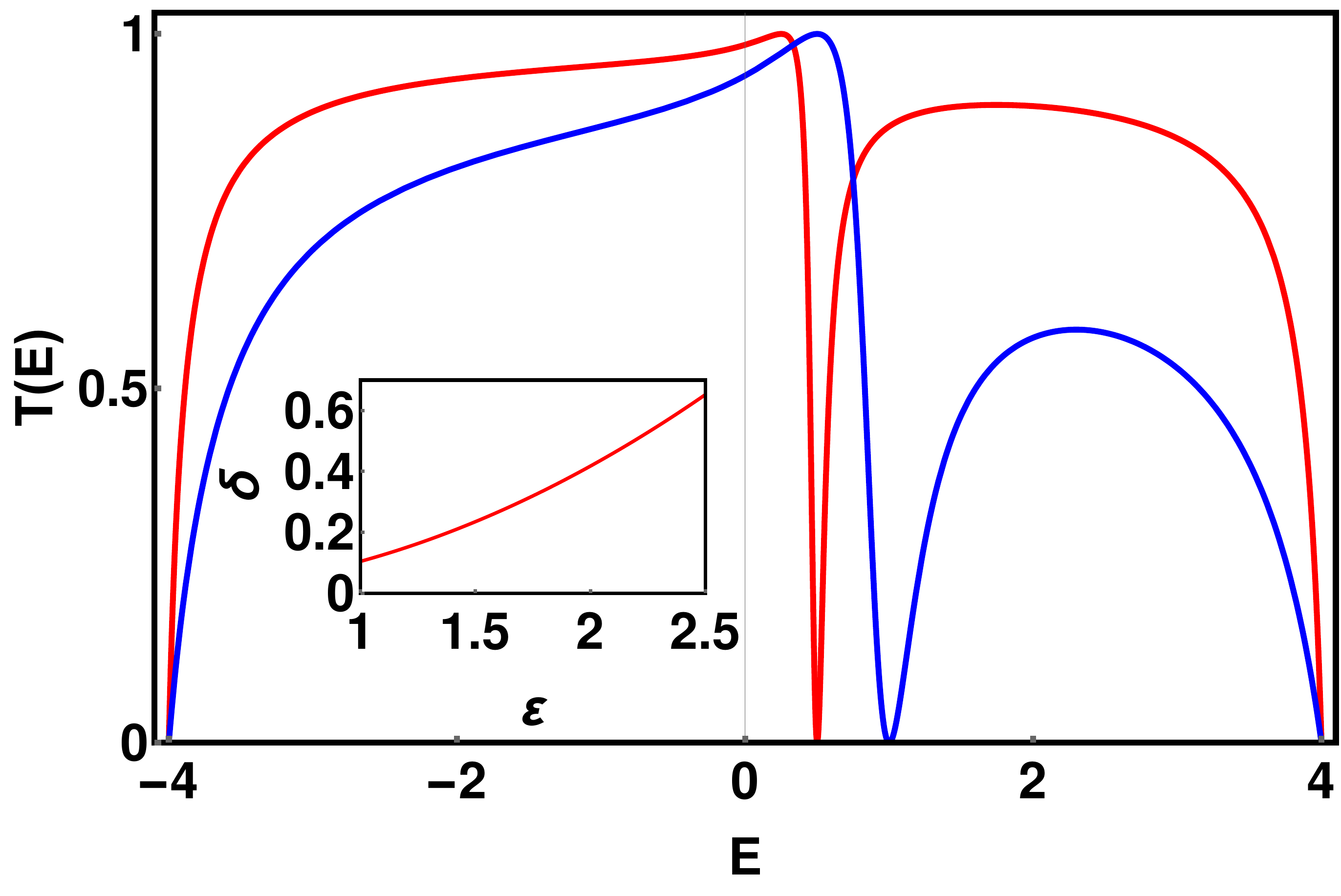}  \\
\end{center}
\caption{Transmission coefficient for different on-site defect potentials ($\epsilon_n^a=1$ (red), $2$ (blue)) for the cross-stitch lattice in the presence of a local on-site defect. 
The inset shows the variation of the width $\delta$ of the resonance 
versus the defect potential $\epsilon=\epsilon_{n_0}^a$. } \label{f2}
\end{figure}
A weak local defect of strength $\epsilon$ will thus lead to a Fano resonance in the dispersive channel. The resonance location is detuned from the original flat band energy by $\epsilon$.
The width of the resonance is quadratic in $\epsilon$. Therefore, the detuning of the resonance off the original flat band energy is well resolved for weak defect strength, and can be used as a detection tool of 
weak imperfections in a flat band lattice.
Our results will hold for any flat band lattice which hosts CLS. The very fact of the existence of a CLS, together with the short rangeness of the tight binding network,
ensures that any local defect will renormalize a CLS, and back-couple it locally into one or several dispersive channels, in the manner of a Fano resonance,
see e.g. \cite{zhu2017fano}.

\section{Disorder}\label{sec:disorder}

Wave propagation in non-periodic media was studied in the seminal work by Anderson in 1958, where absence of diffusion due to uncorrelated disorder has been predicted e.g. in a one-dimensional tight-binding chain \cite{anderson1958Absence}. 
Experimentally, Anderson localization has been observed e.g. with light waves \cite{Lahini2008Anderson}, and Bose-Einsten condensates \cite{billy2008direct,roati2008anderson}. 
Localized states are characterized by an energy-dependent localization length $\xi$ which controls the asymptotic exponential decay of a wavefunction.

Disorder can be interpreted as a finite density of defects inserted into a system. Each individual defect will act as a Fano resonance in the case of a flat band network.  A finite density of such defects
then implies a macroscopic set of Fano resonances - all with slightly detuned (due to disorder) resonance energies. The disorder strength controls both the hybridization of Fano defects with the dispersive
lattice, and the relative detuning of the Fano resonances. The most interesting limit is then the case of weak disorder, where the individual resonance width becomes narrow, while the different
resonances get less detuned and act as a giant macroscopic resonance.

Let us
consider a one-dimensional flat band network (\ref{eq:FB_ham1}) in the presence of a disorder potential in the onsite energy matrix $\epsilon_n$
%
where for each leg of the network $i=1,\dots,\nu$ the onsite energies are uncorrelated random numbers equidistributed over an interval $\epsilon_n^i \in [-W/2,W/2]$. 
The effect of weak uncorrelated disorder potential has been studied in several examples of class $U=1$ and $U=2$ flat band networks
\cite{Leykam2013Flat,flach2014detangling,leykam2017localization}.
In these examples, the scaling law $\xi (W)\sim W^{-\gamma}$ of the localization length $\xi$ as function of the disorder strength as $W\longmapsto 0$ shown surprising exponents $\gamma$ in correspondence of the flat band energy $E_{FB}$, in contrast to the exponent $\gamma=2$ typical of a dispersive band.
In the case of the cross-stitch lattice Eq.(\ref{eq:CS_lin1}), the exponent $\gamma$ of the scaling law of the localization length $\xi \sim W^{-\gamma}$ has been estimated for different values of flat band energy $E_{FB}$, reporting $\gamma=1$ in case of band crossing, $\gamma=1/2$ for the flat band located at the edge of the dispersive band, and a saturation to constant value $\xi \sim c$ for the flat band gapped away from the dispersive one \cite{flach2014detangling}.
Similar exponents have been observed for the diamond chain \cite{Leykam2013Flat} and pyrochlore \cite{leykam2017localization}, as well as class $U=2$ models such as Stub and one-dimensional Lieb lattice \cite{leykam2017localization}.
In higher dimensional lattices, the Fano resonance picture still persists. The computational characterization of eigenstates is 
performed using the participation number $P$ (which counts the number of sites strongly excited in an eigen mode).



In the case of the cross-stitch chain with weak disorder we arrive at a whole array of slightly detuned Fano resonances:
\begin{equation}
(E+\kappa)\, p_n= \left[\epsilon_n^+ + \frac{(\epsilon_n^-)^2}{(E-t) - \epsilon_n^+ 
}\right]\, p_n  - 2\left( p_{n-1} + p_{n+1} \right).
\label{eq:CS-pst}
\end{equation}
The probability distribution function of $z=1/\epsilon_n^+$ is given by 
\[
\mathcal{W}=\frac{2}{z^2}\int \mathcal{P}(y)\mathcal{P}\bigg(\frac{2}{z}-y\bigg) dy
\] 
where $\mathcal{P}(x)$ is the distribution function of $\epsilon_n^+$.  The heavy tails $\mathcal{W}(z) \sim 1/z^2$ result in an effective disorder potential for the dispersive modes which has diverging variance.
This is due to the the slightly detuned CLS Fano resonances acting as a giant strong scattering potential. 
This type of disorder potential has been considered in, for example, the exactly solvable Lloyd model \cite{lloyd1969exactly}, where Thouless \cite{thouless1972relation} and Ishii \cite{ishii1973localization} showed that $\gamma = 1$ within the bulk of the dispersive band and $\gamma=1/2$ at the edge, in contrast with the typical $\gamma=1$ and $\gamma = 2/3$ of the Anderson model. 

Any general disorder potential can be represented as a sum of a symmetric and antisymmetric parts:
  \begin{equation}
\begin{aligned}
\mbox{Symmetric:} \quad \epsilon_n^- = 0 & \Leftrightarrow &  \epsilon_n^a = \epsilon_n^b\;, \\
\mbox{Antisymmetric:} \quad \epsilon_n^+ = 0 & \Leftrightarrow &  \epsilon_n^a = -\epsilon_n^b\;.
\end{aligned}
\label{eq:symAsym}
\end{equation}
\\
The symmetric case $\epsilon_n^- = 0 $ renormalizes the energy of the flat band states, but does not hybridize the CLS with the dispersive states.
The antisymmetric case $\epsilon_n^+ = 0$ instead does not renormalize the CLS energy, but does hybridize them with the dispersive states. 
This latter case is of interest, since it turns the whole set of CLS into one perfect Fano resonance.
For small $\bar{E}=E-E_{FB}$,
the localization length can be obtained as (see \cite{bodyfelt2014flatbands})

\begin{equation}
\xi^{-1} = -2 +  \ln \frac{W^2}{8|\bar{E}|} \ ,\qquad \bar{E}=\pm \frac{\epsilon_0^-\epsilon_{1}^-}{2} \ll \frac{W^2}{4} \;.
\label{loclengthcorrdist=0}
\end{equation} 
\begin{figure}[h]
 \centering
 \includegraphics[ width=0.65\columnwidth]{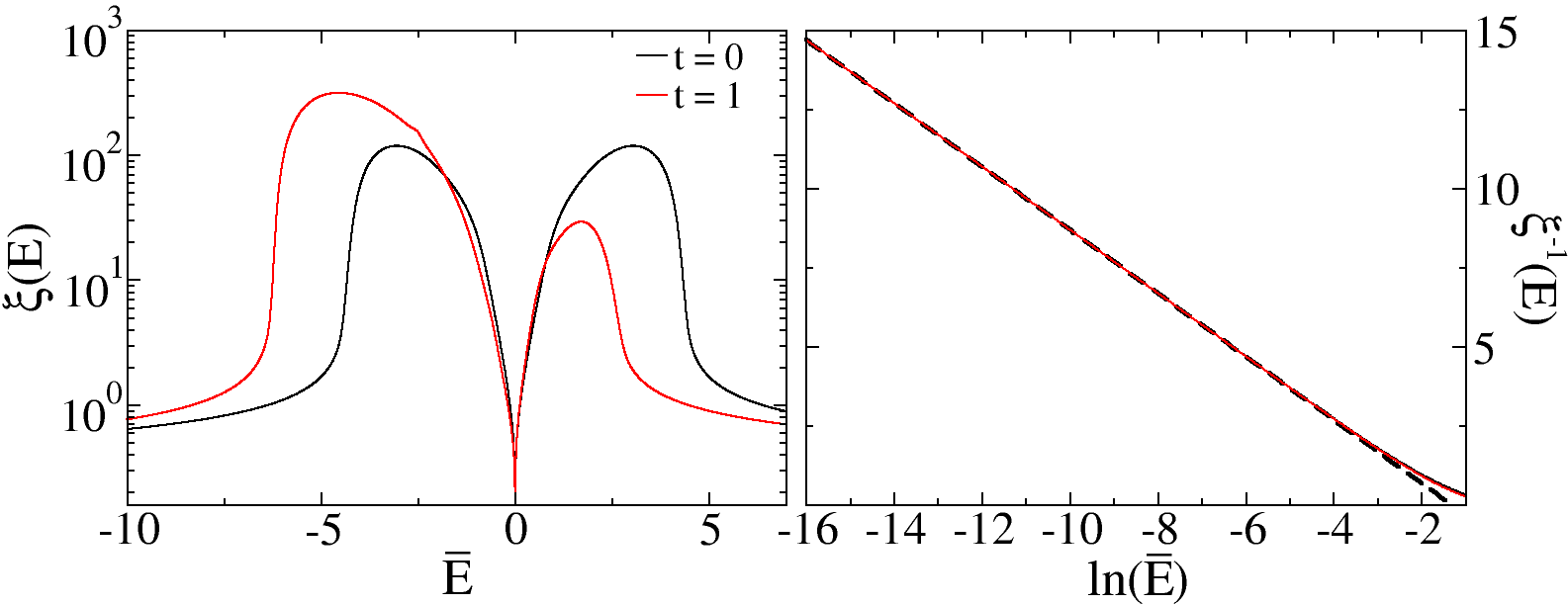}
 \caption{Left plot: Localization length $\xi$ vs $\bar{E}=E-E_{FB}$.
Right plot: $\xi^{-1}$ vs. $\ln \bar{E}$ for $\bar{E} > 0$, same color coding as in left plot.
The dashed line corresponds to Eq.(\ref{loclengthcorrdist=0}). Here, $W=4$ and $t=0$ (black solid) and $\kappa=1$ (red solid). Figure taken from \cite{bodyfelt2014flatbands} }
 \label{fig3_0}
\end{figure}
Irrespective of the strength of the correlated disorder, the localization length vanishes due to resonant scattering as the energy tends towards flat band energy. The numerically calculated 
localization length in Fig. \ref{fig3_0} shows excellent agreement between the numerical data and Eq.(\ref{loclengthcorrdist=0}). Note that  at $E=\kappa$, the equations allow only for a trivial solution $p_n=f_n=0$. All the compact states are hybridized and shifted their energies away, but a significant fraction stays close to flat band energy resulting in divergence of density of states at the flat band energy (not shown here - see \cite{bodyfelt2014flatbands}).



Going from local correlations to global correlations, we can consider the impact of quasiperiodic potentials realized with onsite energies.
In particular for the Aubry-Andr\'e perturbation  
\begin{equation}
\epsilon_n^i= \lambda \cos\left[ 2\pi \left(\alpha n + \theta_i \right) \right]\ ,\qquad i=1,\dots,\nu \;.
\label{eq:AApot_CS}
\end{equation}
Each leg is characterized by the potential strength $\lambda$, and each sequence is offset by all others by the phase shift parameters $\theta_i$.  
The quasiperiodicity is introduced by the incommensurate parameter in the argument of the {\it cosine} functions, which without loss of generality can be set equal $\alpha\in\mathbb{R}\setminus\mathbb{Q}$ in all legs.
The set of CLS will then generate a chain of correlated Fano resonances.



\begin{figure}[h]
 \centering
 \includegraphics[ width=0.4\columnwidth]{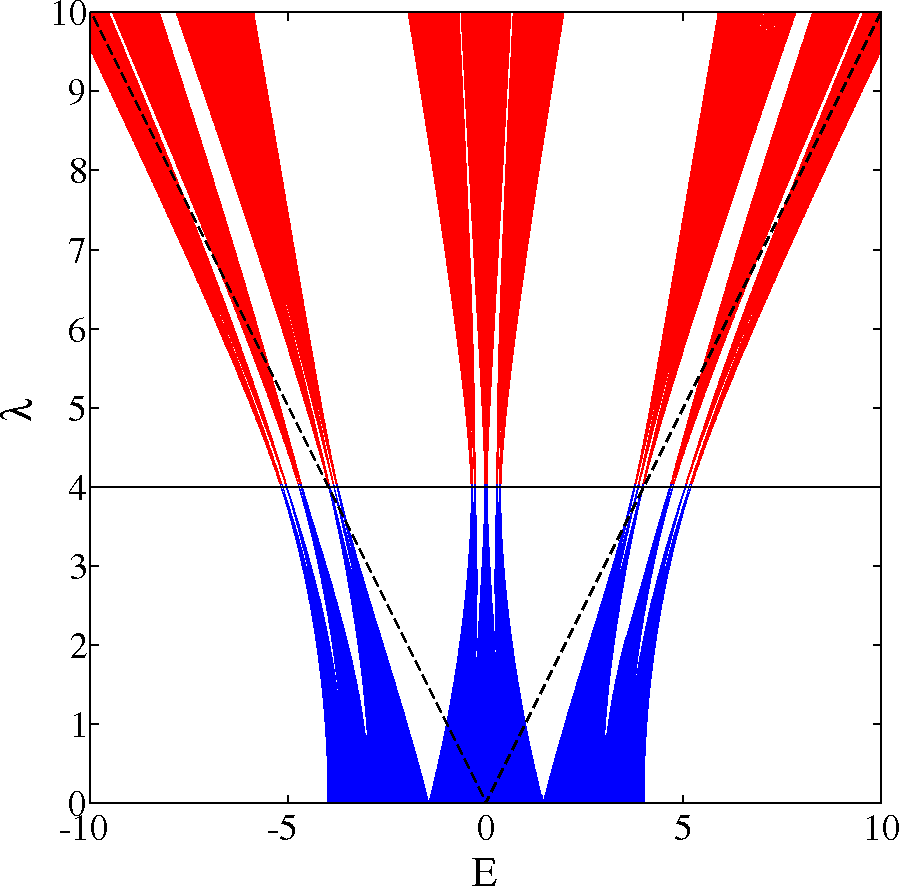}
  \includegraphics[ width=0.4\columnwidth]{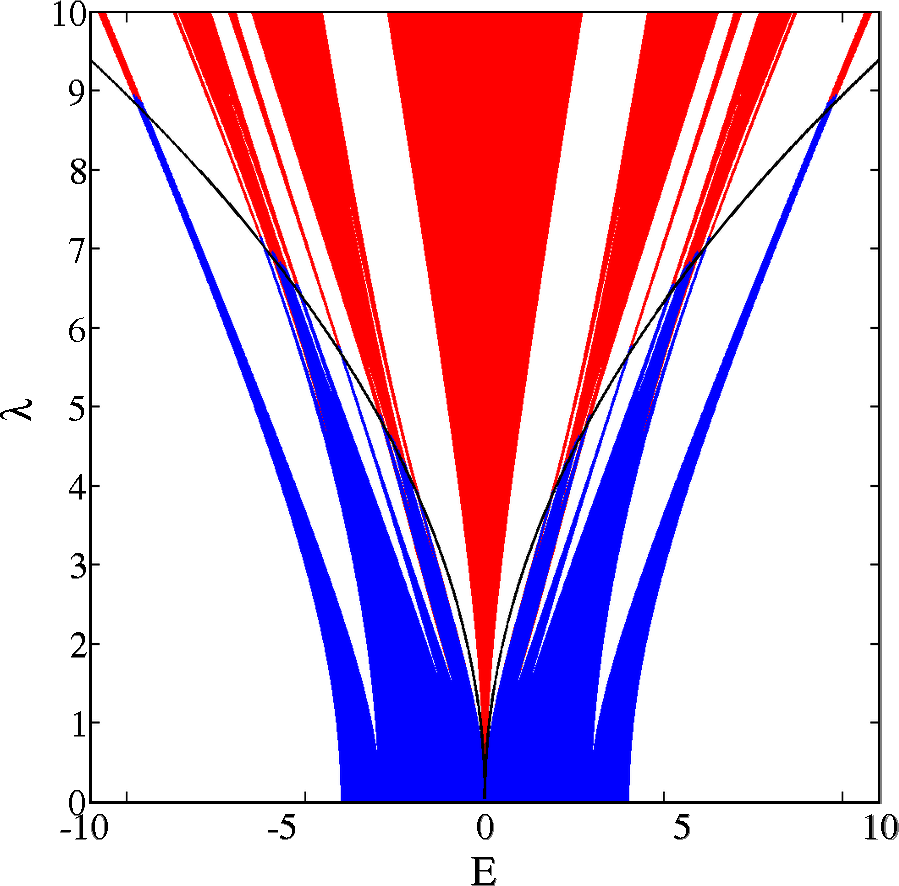}
 \caption{
Spectrum of the cross-stitch lattice. 
Left plot: symmetric case. The dispersive spectrum shows metal-insulator transition at $\lambda_c=4$ (black line).
The Fano state spectrum $\sigma_f$ is omitted (boundaries indicated by black dashed lines). 
Right plot: antisymmetric case. The mobility edge curve corresponds to Eq.(\ref{eq:CS_ME}).
Here: blue = extended states; red: localized states; $\kappa = 0$.
Left figure taken from \cite{bodyfelt2014flatbands}. Right figure taken from \cite{danieli2015flatband}.
}
 \label{fig:QP}
\end{figure}

Let us consider the cross-stitch network Eq.(\ref{eq:CS_lin1}) defined with the Aubry-Andr\'e potential Eq.(\ref{eq:AApot_CS}).
Without loss of generality, we set the phase shift of the $a$-leg to zero $\theta_a=0$.
We will focus on the onsite energy correlations Eq.(\ref{eq:symAsym}), which can be achieved by fine tuning of the phase shift $\theta_b$ on the leg $b$.
The most interesting antisymmetric case $\epsilon_n^+ = 0$ is obtained for 
$\theta_b=1/2$, which reduces Eq.(\ref{eq:CS-pst}) to the following eigenvalue problem
\begin{equation}
(E+\kappa)\, p_n=\dfrac{(\epsilon_n^-)^2}{E - \kappa} \, p_n- 2(p_{n-1} + p_{n+1})\;.
\label{eq:disppor1}
\end{equation}
where
\begin{equation}
(\epsilon_n^-)^2 = \lambda^2\cos^2(2\pi\alpha n) = \dfrac{\lambda^2}{2}\left[1 + \cos(4\pi\alpha n)\right].
\label{eq:bisection}
\end{equation}
Eq.(\ref{eq:disppor1}) then reads as a one-dimensional Aubry-Andr\'e chain with energy-dependent coefficients
\begin{equation}
\begin{split}
&\qquad \tilde{E} \; p_n = \tilde{\lambda}\cos(4\pi\alpha n) - (p_{n-1} + 
p_{n+1}), \\
&\mbox{where }
 \qquad\tilde{E} := \dfrac{E + \kappa}{2} - \dfrac{\lambda^2}{4(E-\kappa)}\ ,\quad \tilde{\lambda} = \dfrac{\lambda^2}{4(E-\kappa)}
 \end{split}
 \label{eq:CS_AA}
\end{equation}
According to \cite{aubry1980analyticity} the transition between the metallic and insulating regimes occurs when $\tilde{\lambda} =2$, which
results in  an analytic expression for the mobility edge, $\lambda_c(E_c)$ :
\begin{equation}
\bigg|\dfrac{\lambda_c^2}{4(E_c-\kappa)}\bigg| = 2 \quad \Rightarrow \quad \lambda_c(E_c)= 
2\sqrt{2|E_c-\kappa|}\;.
\label{eq:CS_ME}
\end{equation}
At the flat band energy $E_{FB}=\kappa $, the mobility edge shows a square root singularity. The coupling of the dispersive states to the giant Fano resonance, generated by the set of CLS with quasiperiodic
modulation of its hybridization strength, carves a tongue of localized states into the metallic regime of the dispersive states, see Fig.\ref{fig:QP}. 
In \cite{danieli2015flatband}, mobility edge transitions between insulating and metallic phase have been obtained for a number of other flat band settings as well.


\section{Nonlinearities}
\label{sec:nonlinear}

It has been rigorously proved that nonlinear lattice wave equations allow for the existence of coherent time-periodic solutions localized in real space called 
{\it discrete breathers}. 
Due to their time periodicity, discrete breathers act as time-periodic scattering potentials for propagating small amplitude waves. 
The confined time periodicity of the scattering potential leads to the existence of several scattering channels, opening the door for destructive interference and
Fano resonances \cite{miroshnichenko2003resonant,kim2000structure,flach2003fano,vicencio2007fano}.

It turns out, that a subclass of flat band lattices with additional nonlinearities  admit {\it compact discrete breathers}, namely solutions of the nonlinear network equations
which are periodic in time and compact in space. These compact breathers are obtained e.g. as the continuation of linear compact localized states. The continuation is accompanied
by a renormalization of the CLS frequency, preserving its compactness.
Consider a one-dimensional flat band network in the presence of Kerr nonlinearity. The model equations read
\begin{equation}
\begin{split}
& i \dot{\vec{\psi}}_n = H_0\vec{\psi}_n + H_1\vec{\psi}_{n+1} + H_1^\dagger\vec{\psi}_{n-1} + \gamma  \mathcal{F}( \vec{\psi}_n) \;, \\
 \mathcal{F}( \vec{\psi}_n) &=  (|\psi_n^1|^2 \psi_n^1,\dots,|\psi_n^\nu|^2\psi_n^\nu)^T =\Bigg[ \sum_{i=1}^\nu |\psi_n^i|^2 {\bf e}_i \otimes {\bf e}_i\Bigg] \vec{\psi}_n \;.
\end{split}
\label{eq:NL_FB_lattice}
\end{equation}
Let us consider the compact localized states Eq.(\ref{eq:FB_states1}) of the linear regime $\gamma = 0$. 
Can this solution persist for nonzero $\gamma \neq 0$ ?
Since the nonlinearity acts locally, zero amplitude sites outside the CLS are not affected.
For the nonzero amplitude sites of the CLS (i.e. for 
$i = 1,\dots ,U$ and  $j =1, \dots, \nu$) we obtain
\begin{equation}
 \Omega A_{i,j} = E_{FB} A_{i,j} + \gamma A_{i,j}^3 \ .
\label{eq:NL_FB4}
\end{equation} 
The nonlinear term yields a shift in the frequency $\Omega = E_{FB} + \gamma A_{i,j}^2$. This expression is equivalent to
\begin{equation}
\forall a_{i,j} \neq  0\quad\Rightarrow\quad  A_{i,j}^2 = \frac{\Omega - E_{FB}}{\gamma} \;.
\label{eq:NL_cls_cond1}
\end{equation}
The linear CLS can be continued as a periodic solution of the nonlinear regime with frequency $\Omega =  E_{FB} + \gamma A^2$ and
compact support if and only if for all the non-zero sites $a_{i,j} \neq  0$ of the linear CLS $|A_{i,j}|^2 \equiv A^2$, where $A$ is some nonzero real number.
Linear CLS that satisfy this condition on their amplitude $A_{i,j}^2 \equiv A^2$ are coined {\it homogeneous} CLS.
{\it Compact discrete breathers} $ \mathcal{C}_{n_0} $ are continued homogeneous CLS of the linear regime with frequency $\Omega = E_{FB} + \gamma A^2$ and $ a_{l,j}  \in \{  0,\pm 1 \}$:
\begin{equation}
 \mathcal{C}_{n_0} (t)= A \Bigg\{ \sum_{l=0}^{U-1} v_l \delta_{n,n_0 + l}\Bigg\} e^{- i \Omega t} \ ,\quad 
 v_l =  \sum_{j=1}^{\nu} a_{l,j}  {\bf e}_j \;.
\label{eq:CLS_stab1}
\end{equation}
The above results  are independent of the number of bands $\nu$ as well as the class  $U$ of the linear  CLS. In the case of the
cross-stitch lattice Eq.(\ref{eq:CS_lin1}), the families of compact discrete breathers are given by 
\begin{equation}
  \mathcal{C}_{n_0} (t) =
   A \left(
\begin{array}{ccc}
1 \\
-1   
\end{array} \right)
\delta_{n,n_0}
  e^{-i \Omega t}  \ .
\label{eq:FB_states1_cs_nl}
\end{equation}
The quantity that parametrizes the families of breathers can be either the renormalized frequency $\Omega $ of the breather amplitude $A$.

In order to study the scattering of the propagation of an extended wave, we consider a small perturbation $\vec{\chi}_n(t)$ of a compact discrete breather $\mathcal{C}_{n_0}$:
\begin{equation}
\vec{\psi}_n (t) =  \mathcal{C}_{n_0}(t) + \vec{\chi}_n(t). 
\label{eq:CB_pert1}
\end{equation}
We linearize Eq.(\ref{eq:NL_FB_lattice}) with respect to $\vec{\chi}_n$ and use $g \equiv \gamma A^2$:
\begin{equation}
i \dot{\vec{\chi}}_n = 
  H_0 \vec{\chi}_n + H_1 \vec{\chi}_{n+1} + H_1^\dagger \vec{\chi}_{n-1} + g \sum_{l=0}^{U-1} \Gamma_l \left(2 \vec{\chi}_{n} + \vec{\chi}_{n} ^* e^{- i 2 \Omega t}   \right) \delta_{n,n_0 + l} \;,
  \label{eq:NL_FB_linearized}
\end{equation}
where the operators 
\begin{equation}
 \Gamma_l = \sum_{j=1}^\nu a_{l,j}^2 {\bf e}_j \otimes {\bf e}_j 
  \label{eq:NL_FB_projectors}
\end{equation}
are the projector operators of a vector on the space of a compact localized state located between the $n_0$-th and the $(n_0+U-1)$-th unit cells.   
The time-dependent linearized equations Eq.(\ref{eq:NL_FB_linearized}) can be mapped to a time-independent eigenvalue problem
\begin{equation}
\vec{\chi}_{n}  = \vec{x}_{n}  e^{- i E t}  +  \vec{y}_{n}^*  e^{- i (2\Omega -  E) t} 
\label{eq:BT}
\end{equation}
where $\vec{x}_n,\vec{y}_n$ are complex vectors of the two scattering channels, $\Omega = E_{FB} + g$ is the frequency of the compact discrete breather, and $ E $ is the frequency
of the propagating wave in the open channel:
\begin{equation}
\begin{split}
\hspace{-3.5mm}
E \vec{x}_n&=   H_0 \vec{x}_n + H_1 \vec{x}_{n+1} + H_1^\dagger \vec{x}_{n-1}  + g \sum_{l=0}^{U-1} \Gamma_l \big(2 \vec{x}_{n} + \vec{y}_{n} \big) \delta_{n,n_0 + l} \;,   \\
(2\Omega - E)  &\vec{y}_n =    H_0 \vec{y}_n + H_1 \vec{y}_{n+1} + H_1^\dagger \vec{y}_{n-1}   + g \sum_{l=0}^{U-1} \Gamma_l \big(2 \vec{y}_{n} + \vec{x}_{n} \big) \delta_{n,n_0 + l}  \;.
\end{split}
\label{eq:NL_FB_lin_rotated}
\end{equation}
The resulting equations describe two independent scattering channels with energy detuning $2\Omega$, and interacting through 
the non-zero amplitude sites of the compact discrete breather $\mathcal{C}_{n_0}$. Here, $\vec{x}_n$ corresponds to the
{\it open} channel, while $\vec{y}_n$ to the {\it closed} channel.
The open channel, away from the alterations induced by the compact breathers located between the $n_0$-th and the $n_0+(U-1)$-th unit cells, support
the spectrum of the linear flat band network. 
Note that the spectra of each of the two channels are composed of several 
bands (flat and dispersive).
 
The linearized system Eq.(\ref{eq:NL_FB_linearized}) in the case of the cross-stitch lattice reads
\vspace{-2mm} 
\begin{equation}
\begin{split}
i\dot{\zeta}_n &= - \zeta_{n-1} - \zeta_{n+1} -  \eta_{n-1} -  \eta_{n+1} - \kappa  \eta_{n} + g \left(2\zeta_{n_0} + e^{-i2\Omega t} \eta_{n_0}^*\right) \delta_{n,n_0} \;,   \\
i\dot{ \eta}_n &= -  \eta_{n-1} -  \eta_{n+1} - \zeta_{n-1} - \zeta_{n+1} - \kappa \zeta_{n} + g \left(2\eta_{n_0} + e^{-i2\Omega t}  \zeta_{n_0}^*\right)  \delta_{n,n_0}  \;.
\end{split}
\label{eq:CS_linearized}
\end{equation}   
Here $\vec{\chi}_n = (\zeta_n , \eta_n)$. The expansion Eq.(\ref{eq:BT}) turns
\begin{equation}
\begin{split}
\zeta_n &= u_n e^{-i Et} +  v_n^* e^{-i(2\Omega - E) t} \;,  \\
\eta_n &=  w_n e^{-i Et} +  z_n^* e^{-i(2\Omega - E) t} \;,
\end{split}
\label{eq:BT_CS}
\end{equation}
and it maps Eq.(\ref{eq:CS_linearized}) into a time-independent problem
\begin{equation}
\begin{split}
E u_n &=  -\big[ u_{n-1} + u_{n+1}  + w_{n-1} + w_{n+1} + h w_{n} \big]  + g ( 2 u_{n_0}+  v_{n_0}) \delta_{n,n_0} \;,   \\
E w_n &=  -\big[ w_{n-1} + w_{n+1}  + u_{n-1} + u_{n+1} + h u_{n} \big]  + g ( 2 w_{n_0}+  z_{n_0}) \delta_{n,n_0} \;,   \\
(2\Omega - E)  v_n &= -  \big[ v_{n-1} + v_{n+1} + z_{n-1} + z_{n+1} + h z_{n}  \big]+ g ( 2 v_{n_0}+  u_{n_0})  \delta_{n,n_0} \;, \\
(2\Omega - E)  z_n &= -  \big[ z_{n-1} + z_{n+1} + v_{n-1} + v_{n+1} + h v_{n}  \big]+ g( 2 z_{n_0}+  w_{n_0})  \delta_{n,n_0} \;.
\end{split}
\label{eq:CS_linerized_Bog1}
\end{equation}
Both the open (first two equations) and the closed (second two equations) of the problem can be further rotated using the coordinate transformation
Eq.(\ref{eq:fano-CS}), detangling the flat band states $f_n^1$ and $f_n^2$ from the dispersive ones $p_n$ and $q_n$ respectively. In particular, also
the compact localized states $f_{n_0}^1$ and $f_{n_0}^2$ located in the $n_0$-th unit cell (where the transversal hopping terms are found) are decoupled
from the correspondent dispersive states  $p_{n_0}$ and $q_{n_0}$, since the alterations and the hopping terms have all the same strength $g\neq 0$.
A combination of the transformation Eq.(\ref{eq:BT}) and the detangling procedure Eq.(\ref{eq:fano-CS}) 
reduces Eq.(\ref{eq:CS_linearized}) to the following one-dimensional equations of the open and the closed channels:
\begin{equation}
\begin{split}
E p_n&= -2(p_{n-1} + p_{n+1}) -  \kappa p_n + g \big[ 2 p_{n_0}+  q_{n_0} \big] \delta_{n,n_0} \;,   \\
(2\Omega - E)  q_n &= -2(q_{n-1}  + q_{n+1} ) -  \kappa q_n + g \big[ 2 q_{n_0} +  p_{n_0} \big] \delta_{n,n_0}  \;.
\end{split}
\label{eq:CS_linearized_rot}
\end{equation}
Let us at first consider the case $g=0$. Indeed, the dispersive bands $E_{1}^O$ and $E_{1}^C$ of the open and the closed channel respectively are
\begin{equation}
\begin{split}
&E_{1}^O =-\kappa + [ -4 , 4 ] \;, \\  
& E_{1}^C = 3\kappa  + [ -4 , 4 ]
\end{split}
\label{eq:closed_channel_bands}
\end{equation}
It follows that for $\kappa\leq 2$ (crossing of the flat band $E_{FB}$ and the dispersive band $E_1$ of the linear cross-stitch lattice),
the dispersive bands overlap $E_{1}^O \cap E_{1}^C \neq \emptyset$. For $\kappa >2$ instead (gapped flat band $E_{FB}$ and
dispersive band $E_1$ of the linear cross-stitch lattice), the dispersive bands of the open and closed channels are disjoint $E_{1}^O \cap E_{1}^C = \emptyset$. 

Next, let us consider Eq.(\ref{eq:CS_linearized_rot}) when both channels are decoupled.
The closed channel admits a localized solution
\begin{equation}
\begin{split}
&q_n = \mathcal{P} x^{|n-n_0|}\ ,\qquad |x|<1,  \quad  \mathcal{P}\neq 0 \;,\\
&\qquad E_L =3\kappa + 2 g -2 \sqrt{4 + g^2} \;.
\end{split}
\label{eq:CS_bound_state}
\end{equation}
The energy $E_L$ of this local mode belongs the interval
$\mathcal{E} \equiv [3\kappa - 4 , 3\kappa]$, where $E_L \rightarrow  3\kappa - 4 $ for $g \rightarrow  0$ and  $E_L \rightarrow  3\kappa  $ for $g \rightarrow  \infty$.
Therefore, the bound states Eq.(\ref{eq:CS_bound_state}) resonate with the dispersive band $E_1^O = [-\kappa+4,-\kappa+4]$  of 
the open channel (i) for any $g>0$ if $|\kappa| < 1$; (ii) only for $g  \leq  \kappa (2-\kappa) / (\kappa-1)$ if $1<\kappa<2$; the (iii) never if $\kappa>2$.

To compute the transmission coefficient $T$ we use the transfer matrix approach discussed and used in \cite{flach2003fano,vicencio2007fano,tong1999wave}. 
Let us define a propagating wave along the $p_n$ chain
\begin{equation}
p_n =
\begin{cases}
& \tau e^{i{\bf k}(n-n_0)} + \sigma e^{- i {\bf k} (n-n_0)} \ ,\quad n<n_0 \\
& \rho e^{i{\bf k}(n-n_0)}\ ,\qquad\qquad\qquad\quad\ \    n\geq n_0
\end{cases}\ ,
\label{eq:CS_ext_wave}
\end{equation}
around the impurity located in the site $n_0$ (we recall $\tau + \sigma = \rho$). 
The transmission coefficient $T({\bf k}) = | \rho / \tau |^2$ follows as
\begin{equation}
T({\bf k})= \frac{16 \sin^2 {\bf k}  }{ 16 \sin^2 {\bf k} +   \Big[ 2g + \frac{ g^2  }{\sqrt{(2\Omega - E_{{\bf k}} + \kappa)^2 - 16} - 2 g  }  \Big]^2 } \;.
\label{eq:NL_CS_TC}
\end{equation}
An additional condition required for the transmission coefficient $T$ 
is that the argument of the square root of the denominator has to be larger than zero for any wave vector {\bf k}. This requirement translates to the inequality $g \geq 4 - \kappa$. 
Then, a Fano resonance occurs when the denominator diverges, or equivalently when $\sqrt{(2\Omega - E_{{\bf k}} + \kappa)^2 - 16} - 2 g = 0$. This condition reads as
\begin{equation}
T({\bf k}) = 0 \quad\Leftrightarrow\quad E_{{\bf k}} = E_L \quad\Leftrightarrow\quad {\bf k} = \arccos \bigg[\frac{\kappa+E_L}{4} \bigg] \;.
\label{eq:NL_CS_fano_res}
\end{equation}
\begin{figure}[h]
 \centering
\includegraphics[width=0.7\linewidth]{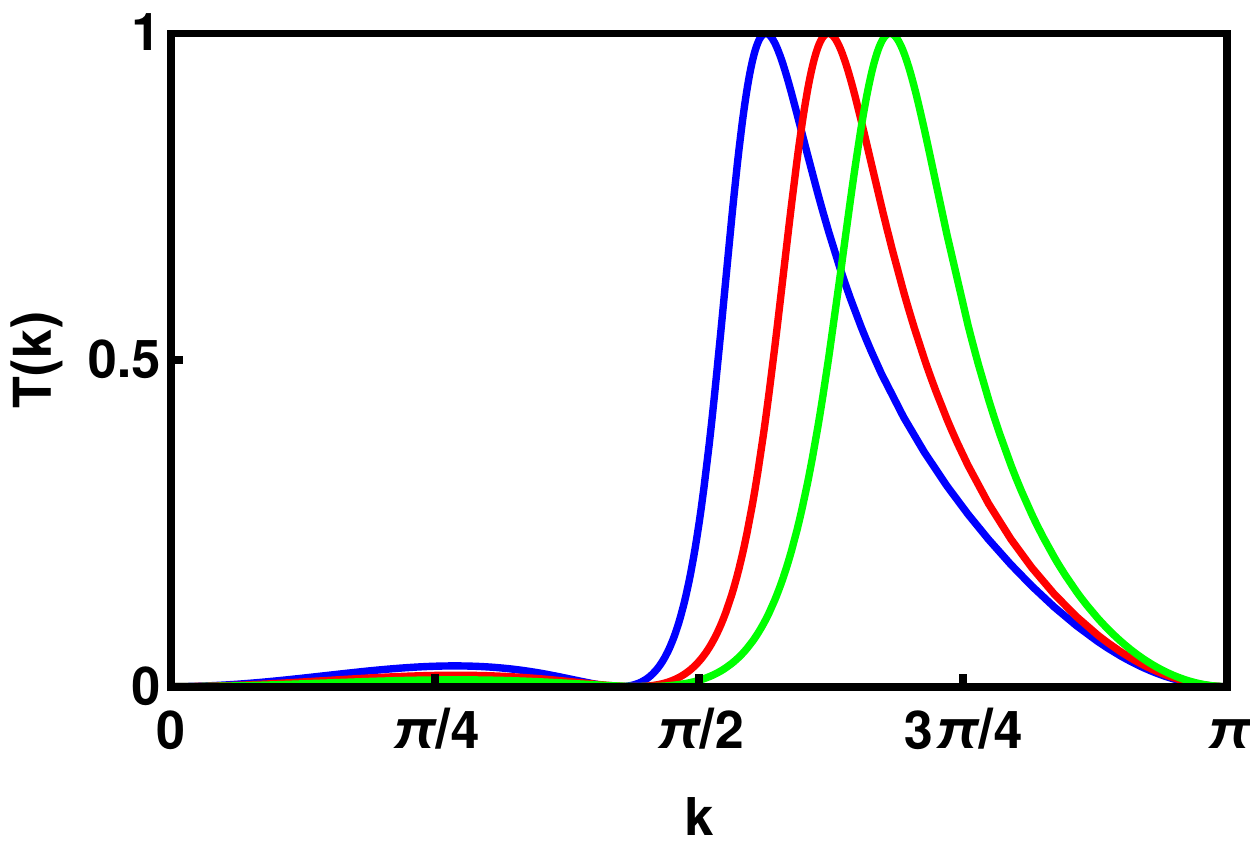}
 \caption{Transmission coefficient $T$ as a function of the momentum ${\bf k}$. Blue $g=4$, Orange $g=5$, Green $g=6$.}
 \label{fig:Fano_NL}
\end{figure}

To summarize, in the nonlinear cross-stitch lattice, a  Fano resonance occurs if and  only if $\kappa < 2$, since otherwise the bound
state of the closed channel Eq.(\ref{eq:CS_bound_state}) is out of resonance with the dispersive band of the open channel. Furthermore, the condition Eq.(\ref{eq:NL_CS_fano_res})
is subject to the requirement $g \geq 4 - \kappa$, which translates to $4 - \kappa \leq g  \leq  \kappa (2-\kappa) / (\kappa-1)$, if $1<\kappa<2$. 
In Fig.\ref{fig:Fano_NL} we show the transmission coefficient $T$ as function of the wave vector {\bf k} in the case of $\kappa = 0$ for different values of $g\geq 4$. Indeed,
zeros of the curves appear, indicating absence of transmission of a propagating wave.

\section{Conclusions}

In this chapter, we have discussed phenomena of total reflection of propagating waves in flat band networks, due to impurities,
disorder and quasiperiodic potentials, and to compact discrete breathers induced by the presence of nonlinearities.
These phenomena have been outlined generally for classes of flat band lattices, and they have been analyzed in detail
in the case of the cross-stitch lattice. The flat band lattices are interesting candidates 
to visualize Fano resonances because of the presence of compact localized states which serve as Fano states in the presence 
of proper perturbations.
The defect-induced Fano resonance 
has been shown to result in complete suppression of the propagation. Many defects in lattice can result in distinctive 
transmission profile characteristics and a macroscopic number of defects takes the role of a disorder which, either correlated or uncorrelated, has a profound
effect on the localization of dispersive states 
with anomalous scaling properties. 
Experimentally, flat band network Fano resonances can be studied in photonic crystals, optical lattices or even electronic circuits. 
Further, flat band networks and their Fano resonances can be used to
engineer different types of spectral singularities or mobility edges in lattice systems and to control wave transport.
In the presence of nonlinearities, novel compact discrete breather solutions turn into tunable Fano resonance scatterers.


\section*{Acknowldegment}
This work was supported by the Institute for Basic Science, Project Code (IBS-R024-D1)

 \bibliographystyle{spphys}
 \bibliography{flatband}

\begin{thebibliography}{10}
\providecommand{\url}[1]{{#1}}
\providecommand{\urlprefix}{URL }
\expandafter\ifx\csname urlstyle\endcsname\relax
  \providecommand{\doi}[1]{DOI \discretionary{}{}{}#1}\else
  \providecommand{\doi}{DOI \discretionary{}{}{}\begingroup
  \urlstyle{rm}\Url}\fi

\bibitem{fano1935sullo}
U.~Fano, \emph{Sullo spettro di assorbimento dei gas nobili presso il limite
  dello spettro d'arco}, Il Nuovo Cimento \textbf{12}, 154
\newblock
\newblock  (1935)

\bibitem{fano1961effects}
U.~Fano, \emph{Effects of configuration interaction on intensities and phase
  shifts}, Phys. Rev. \textbf{124}, 1866
\newblock
\newblock  (1961)

\bibitem{miroshnichenko2010fano}
A.E. Miroshnichenko, S.~Flach, Y.S. Kivshar, \emph{Fano resonances in nanoscale
  structures}, Rev. of Mod. Phys. \textbf{82}, 2257
\newblock
\newblock  (2010)

\bibitem{johnson2005charge}
A.~Johnson, C.~Marcus, M.~Hanson, A.~Gossard, \emph{Charge sensing of excited
  states in an isolated double quantum dot}, Phys. Rev. B \textbf{71}, 115333
\newblock
\newblock  (2005)

\bibitem{gores2000fano}
J.~G{\"o}res, D.~Goldhaber-Gordon, S.~Heemeyer, M.~Kastner, H.~Shtrikman,
  D.~Mahalu, U.~Meirav, \emph{Fano resonances in electronic transport through a
  single-electron transistor}, Phys. Rev. B \textbf{62}, 2188
\newblock
\newblock  (2000)

\bibitem{bulka2001fano}
B.R. Bu{\l}ka, P.~Stefa{\'n}ski, \emph{Fano and Kondo resonance in electronic
  current through nanodevices}, Phys. Rev. Lett. \textbf{86}, 5128
\newblock
\newblock  (2001)

\bibitem{torio2004spin}
M.~Torio, K.~Hallberg, S.~Flach, A.~Miroshnichenko, M.~Titov, \emph{Spin
  filters with Fano dots}, Eur. Phys. J. B \textbf{37}, 399
\newblock
\newblock  (2004)

\bibitem{franco2003fano}
R.~Franco, M.~Figueira, E.~Anda, \emph{Fano resonance in electronic transport
  through a quantum wire with a side-coupled quantum dot: X-boson treatment},
  Phys. Rev. B \textbf{67}, 155301
\newblock
\newblock  (2003)

\bibitem{rybin2010bragg}
M.~Rybin, A.~Khanikaev, M.~Inoue, A.~Samusev, M.~Steel, G.~Yushin, M.~Limonov,
  \emph{Bragg scattering induces Fano resonance in photonic crystals},
  Photonics Nanostruct. \textbf{8}, 86
\newblock
\newblock  (2010)

\bibitem{rybin2009fano}
M.~Rybin, A.~Khanikaev, M.~Inoue, K.~Samusev, M.~Steel, G.~Yushin, M.~Limonov,
  \emph{Fano resonance between Mie and Bragg scattering in photonic crystals},
  Phys. Rev. Lett. \textbf{103}, 023901
\newblock
\newblock  (2009)

\bibitem{mukhopadhyay2011signature}
S.~Mukhopadhyay, R.~Biswas, C.~Sinha, \emph{Signature of quantum interference
  and the Fano resonances in the transmission spectrum of bilayer graphene
  nanostructure}, J. of Appl. Phys. \textbf{110}, 014306
\newblock
\newblock  (2011)

\bibitem{tong1999wave}
P.~Tong, B.~Li, B.~Hu, \emph{Wave transmission, phonon localization, and heat
  conduction of a one-dimensional Frenkel-Kontorova chain}, Phys. Rev. B
  \textbf{59}, 8639
\newblock
\newblock  (1999)

\bibitem{flach2014detangling}
S.~Flach, D.~Leykam, J.D. Bodyfelt, P.~Matthies, A.S. Desyatnikov,
  \emph{Detangling flat bands into Fano lattices}, EPL (Europhysics Letters)
  \textbf{105}, 30001
\newblock
\newblock  (2014)

\bibitem{perchikov2017flat}

\newblock
\newblock  (2017)

\bibitem{huber2010bose}
S.D. Huber, E.~Altman, \emph{Bose condensation in flat bands}, Phys. Rev. B
  \textbf{82}, 184502
\newblock
\newblock  (2010)

\bibitem{aoki1996hofstadter}
H.~Aoki, M.~Ando, H.~Matsumura, \emph{Hofstadter butterflies for flat bands},
  Phys. Rev. B \textbf{54}, R17296
\newblock
\newblock  (1996)

\bibitem{Leykam2013Flat}
D.~Leykam, S.~Flach, O.~Bahat-Treidel, A.S. Desyatnikov, \emph{Flat band
  states: Disorder and nonlinearity}, Phys. Rev. B \textbf{88}, 224203
\newblock
\newblock  (2013)

\bibitem{leykam2017localization}
D.~Leykam, J.D. Bodyfelt, A.S. Desyatnikov, S.~Flach, \emph{Localization of
  weakly disordered flat band states}, Eur. Phys. J. B \textbf{90}, 1
\newblock
\newblock  (2017)

\bibitem{bodyfelt2014flatbands}
J.D. Bodyfelt, D.~Leykam, C.~Danieli, X.~Yu, S.~Flach, \emph{Flatbands under
  Correlated Perturbations}, Phys. Rev. Lett. \textbf{113}, 236403
\newblock
\newblock  (2014)

\bibitem{danieli2015flatband}
C.~Danieli, J.D. Bodyfelt, S.~Flach, \emph{Flat-band engineering of mobility
  edges}, Phys. Rev. B \textbf{91}, 235134
\newblock
\newblock  (2015)

\bibitem{vicencio2015observation}
R.A. Vicencio, C.~Cantillano, L.~Morales-Inostroza, B.~Real,
  C.~Mej\'{\i}a-Cort\'es, S.~Weimann, A.~Szameit, M.I. Molina,
  \emph{Observation of Localized States in Lieb Photonic Lattices}, Phys. Rev.
  Lett. \textbf{114}, 245503
\newblock
\newblock  (2015)

\bibitem{weimann2016transport}
S.~Weimann, L.~Morales-Inostroza, B.~Real, C.~Cantillano, A.~Szameit, R.A.
  Vicencio, \emph{Transport in Sawtooth photonic lattices}, Optics letters
  \textbf{41}, 2414
\newblock
\newblock  (2016)

\bibitem{brandes2005coherent}
T.~Brandes, \emph{Coherent and collective quantum optical effects in mesoscopic
  systems}, Phys. Rep. \textbf{408}, 315
\newblock
\newblock  (2005)

\bibitem{taie2015coherent}
S.~Taie, H.~Ozawa, T.~Ichinose, T.~Nishio, S.~Nakajima, Y.~Takahashi,
  \emph{Coherent driving and freezing of bosonic matter wave in an optical Lieb
  lattice}, Sci. Adv. \textbf{1}
\newblock
\newblock  (2015)

\bibitem{bellec2013tight}
M.~Bellec, U.~Kuhl, G.~Montambaux, F.~Mortessagne, \emph{Tight-binding
  couplings in microwave artificial graphene}, Phys. Rev. B \textbf{88}, 115437
\newblock
\newblock  (2013)

\bibitem{casteels2016probing}
W.~Casteels, R.~Rota, F.~Storme, C.~Ciuti, \emph{Probing photon correlations in
  the dark sites of geometrically frustrated cavity lattices}, Phys. Rev. A
  \textbf{93}, 043833
\newblock
\newblock  (2016)

\bibitem{masumoto2012exciton}
N.~Masumoto, N.Y. Kim, T.~Byrnes, K.~Kusudo, A.~L{\"o}ffler, S.~H{\"o}fling,
  A.~Forchel, Y.~Yamamoto, \emph{Exciton--polariton condensates with flat bands
  in a two-dimensional kagome lattice}, New J. Phys. \textbf{14}, 065002
\newblock
\newblock  (2012)

\bibitem{qiu2016designing}
W.X. Qiu, S.~Li, J.H. Gao, Y.~Zhou, F.C. Zhang, \emph{Designing an artificial
  Lieb lattice on a metal surface}, Phys. Rev. B \textbf{94}, 241409
\newblock
\newblock  (2016)

\bibitem{mielke1991ferromagnetism}
A.~Mielke, \emph{Ferromagnetism in the Hubbard model on line graphs and further
  considerations}, J. Phys. A: Math. Gen. \textbf{24}, 3311
\newblock
\newblock  (1991)

\bibitem{tasaki1992ferromagnetism}
H.~Tasaki, \emph{Ferromagnetism in the Hubbard models with degenerate
  single-electron ground states}, Phys. Rev. Lett. \textbf{69}, 1608
\newblock
\newblock  (1992)

\bibitem{dias2015origami}
R.G. Dias, G.J. D., \emph{Origami rules for the construction of localized
  eigenstates of the Hubbard model in decorated lattices}, Sci. Rep.
  \textbf{5}, 16852
\newblock
\newblock  (2015)

\bibitem{morales2016simple}
L.~Morales-Inostroza, R.A. Vicencio, \emph{Simple method to construct flat-band
  lattices}, Phys. Rev. A \textbf{94}, 043831
\newblock
\newblock  (2016)

\bibitem{maimaiti2017compact}
W.~Maimaiti, A.~Andreanov, H.C. Park, O.~Gendelman, S.~Flach, \emph{Compact
  localized states and flat-band generators in one dimension}, Phys. Rev. B
  \textbf{95}, 115135
\newblock
\newblock  (2017)

\bibitem{Ramachandran2017chiral}
A.~Ramachandran, A.~Andreanov, S.~Flach, \emph{Chiral flat bands: Existence,
  engineering, and stability}, Phys. Rev. B \textbf{96}, 161104
\newblock
\newblock  (2017)

\bibitem{vidal2000interaction}
J.~Vidal, B.~Dou{\c{c}}ot, R.~Mosseri, P.~Butaud, \emph{Interaction induced
  delocalization for two particles in a periodic potential}, Phys. Rev. Lett.
  \textbf{85}, 3906
\newblock
\newblock  (2000)

\bibitem{Leykam2017Flat}
D.~Leykam, S.~Flach, Y.D. Chong, \emph{Flat bands in lattices with
  non-Hermitian coupling}, Phys. Rev. B \textbf{96}, 064305
\newblock
\newblock  (2017)

\bibitem{khomeriki2016landau}
R.~Khomeriki, S.~Flach, \emph{Landau-Zener Bloch Oscillations with Perturbed
  Flat Bands}, Phys. Rev. Lett. \textbf{116}, 245301
\newblock
\newblock  (2016)

\bibitem{kolovsky2017topological}
A.R. Kolovsky, A.~Ramachandran, S.~Flach, \emph{Topological flat Wannier-Stark
  bands}, arXiv preprint arXiv:1707.05953
\newblock
\newblock  (2017)

\bibitem{peotta2015superfluidity}
S.~Peotta, P.~T{\"{o}}rm{\"{a}}, \emph{Superfluidity in topologically
  nontrivial flat bands}, Nat. Comm. \textbf{6}, 8944
\newblock
\newblock  (2015)

\bibitem{zhu2017fano}
R.~Zhu, C.~Cai, \emph{Fano resonance via quasibound states in time-dependent
  three-band pseudospin-1 Dirac-Weyl systems}, J. Appl. Phys. \textbf{122},
  124302
\newblock
\newblock  (2017)

\bibitem{anderson1958Absence}
P.W. Anderson, \emph{Absence of Diffusion in Certain Random Lattices}, Phys.
  Rev. \textbf{109}, 1492
\newblock
\newblock  (1958)

\bibitem{Lahini2008Anderson}
Y.~Lahini, A.~Avidan, F.~Pozzi, M.~Sorel, R.~Morandotti, D.N. Christodoulides,
  Y.~Silberberg, \emph{Anderson localization and nonlinearity in
  one-dimensional disordered photonic lattices}, Phys. Rev. Lett. \textbf{100},
  013906
\newblock
\newblock  (2008)

\bibitem{billy2008direct}
J.~Billy, V.~Josse, Z.~Zuo, A.~Bernard, B.~Hambrecht, P.~Lugan, D.~Cl{\'e}ment,
  L.~Sanchez-Palencia, P.~Bouyer, A.~Aspect, \emph{Direct observation of
  Anderson localization of matter waves in a controlled disorder}, Nature
  \textbf{453}, 891
\newblock
\newblock  (2008)

\bibitem{roati2008anderson}
G.~Roati, C.~D'errico, L.~Fallani, M.~Fattori, C.~Fort, M.~Zaccanti,
  G.~Modugno, M.~Modugno, M.~Inguscio, in \emph{Pushing The Frontiers Of Atomic
  Physics} (World Scientific, 2009),
\newblock
\newblock pp. 190--200

\bibitem{lloyd1969exactly}
P.~Lloyd, \emph{Exactly solvable model of electronic states in a
  three-dimensional disordered Hamiltonian: non-existence of localized states},
  J. of Phys. C \textbf{2}, 1717
\newblock
\newblock  (1969)

\bibitem{thouless1972relation}
D.~Thouless, \emph{A relation between the density of states and range of
  localization for one dimensional random systems}, J. Phys. C \textbf{5}, 77
\newblock
\newblock  (1972)

\bibitem{ishii1973localization}
K.~Ishii, \emph{Localization of eigenstates and transport phenomena in the
  one-dimensional disordered system}, Progress of Theoretical Physics
  Supplement \textbf{53}, 77
\newblock
\newblock  (1973)

\bibitem{aubry1980analyticity}
S.~Aubry, G.~Andr{\'e}, \emph{Analyticity breaking and Anderson localization in
  incommensurate lattices}, Ann. Israel Phys. Soc \textbf{3}, 18
\newblock
\newblock  (1980)

\bibitem{miroshnichenko2003resonant}
A.~Miroshnichenko, S.~Flach, B.~Malomed, \emph{Resonant scattering of
  solitons}, Chaos \textbf{13}, 874
\newblock
\newblock  (2003)

\bibitem{kim2000structure}
S.W. Kim, S.~Kim, \emph{The structure of eigenmodes and phonon scattering by
  discrete breathers in the discrete nonlinear Schr{\"o}dinger chain}, Physica
  D \textbf{141}, 91
\newblock
\newblock  (2000)

\bibitem{flach2003fano}
S.~Flach, A.~Miroshnichenko, V.~Fleurov, M.~Fistul, \emph{Fano resonances with
  discrete breathers}, Phys. Rev. Lett. \textbf{90}, 084101
\newblock
\newblock  (2003)

\bibitem{vicencio2007fano}
R.A. Vicencio, J.~Brand, S.~Flach, \emph{Fano blockade by a Bose-Einstein
  condensate in an optical lattice}, Phys. Rev. Lett. \textbf{98}, 184102
\newblock
\newblock  (2007)

\end{thebibliography}

\printindex
\end{document}